\documentclass[twoside,11pt]{article}
\usepackage{graphicx} 
\newcommand{\MN}{M_1}
\newcommand{\epsN}{\varepsilon_1}
\newcommand{\lambdaN}{\lambda_1}
\newcommand{\mNN}{\widetilde{m}_1}
\newcommand{\mNNstar}{\widetilde{m}^*_1}
\newcommand{\Tsph}{10^{11\div 12}}

\newcommand{\NR}{N}
\newcommand{\higgs}{h}
\newcommand{\lepton}{\ell}
\newcommand{\ratio}[1]{\frac{\Delta \Gamma_{#1}}{\Delta \Gamma}}
\renewcommand{\ratio}[1]{c_{#1}}
\newcommand{\riga}[1]{\noalign{\hbox{\parbox{\textwidth}{#1}}}\nonumber}

\newcommand{\One}{\hbox{1\kern-.24em I}}

\newcommand{\GeV}{\,{\rm GeV}}
\newcommand{\TeV}{\,{\rm TeV}}

\newcommand{\eV}{\,{\rm eV}}
\newcommand{\NP}{Nucl. Phys.}
\newcommand{\JHEP}{J.HEP}
\newcommand{\PRL}{Phys. Rev. Lett.}
\newcommand{\PL}{Phys. Lett.}
\newcommand{\PR}{Phys. Rev.}

\newcommand{\eq}[1]{~{\rm (\ref{eq:#1})}}
\newcommand{\sys}[1]{~{\rm (\ref{sys:#1})}}
\newcommand{\fig}[1]{~{\rm \ref{fig:#1}}}
\newcommand{\md}[1]{\langle#1\rangle}

\newcommand{\K}{\hbox{K}}

\newcommand{\be}{\begin{equation}}
\newcommand{\ee}{\end{equation}}
\newcommand{\ba}{\begin{array}}
\newcommand{\ea}{\end{array}}

\newcommand{\mysection}[1]{\section{\Blue #1\Black }\setcounter{equation}{0}}

\newcommand{\MGUT}{M_{\rm G}}
\def\Red{}
\def\Black{}
\def\Blue{}

\newcommand{\lascia}[1]{}
\makeatletter
\def\art{\@ifnextchar[{\eart}{\oart}}
\def\eart[#1]#2#3#4#5#6{{\rm #2}, {\em #3 \bf #4} {\rm (#6) #5} ({\em #1})}
\def\hepart[#1]#2{{\rm #2, \em#1}}
\newcommand{\oart}[5]{{\rm #1}, {\em #2 \bf #3} {\rm (#5) #4}}

\newcounter{alphaequation}[equation]
\def\thealphaequation{\theequation\hbox to
0.6em{\hfil\alph{alphaequation}\hfil}}
\def\eqnsystem#1{
\def\@eqnnum{{\rm (\thealphaequation)}}
\def\@@eqncr{\let\@tempa\relax \ifcase\@eqcnt \def\@tempa{& & &} \or
  \def\@tempa{& &}\or \def\@tempa{&}\fi\@tempa
  \if@eqnsw\@eqnnum\refstepcounter{alphaequation}\fi
\global\@eqnswtrue\global\@eqcnt=0\cr}
\refstepcounter{equation} \let\@currentlabel\theequation \def\@tempb{#1}
\ifx\@tempb\empty\else\label{#1}\fi
\refstepcounter{alphaequation}
\let\@currentlabel\thealphaequation
\global\@eqnswtrue\global\@eqcnt=0 \tabskip\@centering\let\\=\@eqncr
$$\halign to \displaywidth\bgroup \@eqnsel\hskip\@centering
$\displaystyle\tabskip\z@{##}$&\global\@eqcnt\@ne
\hskip2\arraycolsep\hfil${##}$\hfil& \global\@eqcnt\tw@\hskip2\arraycolsep
$\displaystyle\tabskip\z@{##}$\hfil
\tabskip\@centering&\llap{##}\tabskip\z@\cr}

\def\endeqnsystem{\@@eqncr\egroup$$\global\@ignoretrue} \makeatother

\def\diag{\mathop{\rm diag}}

\def\Tr{\mathop{\rm Tr}}
\def\circa#1{\,\raise.3ex\hbox{$#1$\kern-.75em\lower1ex\hbox{$\sim$}}\,}
\begin{document}
\begin{quote}
{\em 11/11/1999}\hfill {\bf SNS-PH/99-15}\\
hep-ph/9911315 \hfill{\bf IFUP-TH/99-55}
\end{quote}
\bigskip\bigskip
\centerline{\LARGE\bf\Red Baryogenesis through leptogenesis}
\bigskip\bigskip\Black
\centerline{\large\bf Riccardo Barbieri$^{\rm ab}$, Paolo Creminelli$^{\rm a}$,} \vspace{0.2cm}
\centerline{\large\bf Alessandro Strumia$^{\rm bc}$ {\rm and} Nikolaos Tetradis$^{\rm abd}$} \vspace{0.3cm}

\bigskip
\centerline{(a) \em Scuola Normale Superiore, Piazza dei Cavalieri 7, I-56126 Pisa, Italy}
\centerline{(b) \em INFN, sezione di Pisa, I-56127 Pisa, Italia}
\centerline{(c) \em Dipartimento di fisica, Universit\`a di Pisa, I-56127 Pisa, Italia}
\centerline{(d) \em Department of Physics, University of Athens, GR-15771 Athens, Greece}

\bigskip\bigskip\Blue
\centerline{\large\bf Abstract}
\begin{quote}\large\indent
Baryogenesis by heavy-neutrino decay and sphaleron reprocessing of both baryon and lepton number is reconsidered,
paying special attention to the flavour structure of the general evolution equations and developing an
approximate but sufficiently accurate analytic solution to the prototype evolution equation.
Two different models of neutrino masses are examined, based on an Abelian U(1) or a non-Abelian U(2)
family symmetry. We show that a consistent picture of baryogenesis can emerge in both cases,
although with significant differences.

\end{quote}\Black

\setcounter{equation}{0}
\renewcommand{\theequation}{\thesection.\arabic{equation}}

\noindent

\section*{\Blue Introduction to the revised version\Black}
This paper contained two different  original results:
\begin{itemize}
\item[1] We  discussed how to include flavour effects in computations of thermal leptogenesis,
showing that one must write Boltzmann equations for a $3\times 3$ density matrix
that describes how the lepton asymmetry is shared between the 3 flavours.

\item[2]  We gave approximated semi-analytical solutions to the standard Boltzmann equations
and presented numerical results in a way that covers all the parameter space systematically.
\end{itemize}
Part 1 remains unchanged.
Part 2 has been revised using the corrected Boltzmann equations of~\cite{thermal}.
We do not list the changes we made.
Final expressions are now simpler.


\mysection{Introduction}
The need for a mechanism to generate the baryon asymmetry of the universe after 
an early era of cosmological inflation is one of the main indications for physics
beyond the Standard Model (SM). Several such mechanisms have been proposed, 
such as baryon-number violating interactions at tree level in Grand Unified Theories
(GUTs), or sphaleron transitions taking place at the quantum level in electroweak
baryogenesis.

An even stronger indication for physics beyond the SM is provided nowadays by 
the anomalies in neutrino physics~\cite{sun-exp,atm-exp,LSND}, which can be interpreted in terms of neutrino 
oscillations. Using a simple-minded {\em see-saw} formula, \hbox{$m_\nu = (200 \GeV)^2/M$},
a neutrino mass between $0.03 \eV$ and a few eVs, as seemingly implied by a 
coherent interpretation of the various neutrino results, suggests a mass scale
for new physics between $10^{15}$ and $10^{13} \GeV$. 
In turn, this mass scale could be related to the lepton number violating exchange 
of heavy right-handed Majorana neutrinos $\NR$.

These recent findings in neutrino physics have stimulated the reconsideration of
the possibility that the baryon asymmetry may be generated by a lepton asymmetry,
arising from the out-of-equilibrium decay of the heavy $\NR$~\cite{original}. This conversion
of asymmetries would occur by the reprocessing of both baryon ($B$) and lepton
number ($L$) in sphaleron transitions. Ideally, from a detailed model of neutrino 
masses, both light and heavy, one would like to compute the difference between 
the number density of baryons and antibaryons, normalized to the entropy density
of the universe, $(n_B-n_{\bar B})/s \equiv Y_B$, at the time of nucleosynthesis, known to be 
in the $10^{-(10\div 11)}$ range from the primordial abundances of the light
elements~\cite{nucleos}.
In order to minimize uncertainties, the reheating temperature after inflation $T_{\rm rh}$ is assumed to be 
bigger than the mass of the decaying neutrino~\cite{M>Trh}.

With this general programme in mind, in this paper we add a few elements to the
standard analysis of baryogenesis via leptogenesis: the consideration of the flavour
structure of the problem and an approximate but sufficiently accurate analytic
solution of the relevant evolution equations. These results would be fully relevant
if a sufficiently detailed model of neutrino masses existed, which is not the case
at present. Nevertheless, we apply our considerations to two different models
of neutrino masses, based on an Abelian U(1) or a non-Abelian U(2) family symmetry respectively. 
A consistent picture of baryogenesis can emerge in both cases,
although with significant differences, thus strengthening the view that makes leptogenesis an appealing mechanism for baryogenesis.

\mysection{General setting of the problem}\label{setting}
Sphaleron transitions, in equilibrium at temperatures above about $100
\GeV$, violate $B$ and $L$ while conserving the quantities $\Delta_i \equiv \frac{1}{3} B - L_i$,
$i = e, \mu, \tau$~\cite{sphaleronOld,sphaleronNew}. As a consequence, by imposing the equilibrium conditions in the 
SM on all chemical potentials, one obtains
below the electroweak phase transition~\cite{mucalcoli}
\be
\label{conversion}\label{Deltai}
Y_B = \frac{12}{37} \sum_i \left.Y_{\Delta_i}\right|_0,\qquad
Y_{\Delta_i} \equiv \frac{1}{3} Y_B - Y_{L_i},
\ee
which specifies $Y_B$ in terms of the asymmetries $Y_{\Delta_i}|_0$, either initial or generated before the electroweak
phase transition.

In the pure SM $\Delta_i$ are conserved.
Similarly $\sum_i Y_{\Delta_i}$ cannot be generated in a minimal SU(5) model.
In order to generate a baryon 
asymmetry, we have to assume that an out-of-equilibrium interaction around a temperature $T^*$ violates
(some of) the $\Delta_i$ and produces a net $Y_{\Delta_i}|_0$ different from zero: $\NR$ decay is the example
that is of interest here, in which case $T^*$ is of order of the
$\NR$ mass, $\MN $. This decay acts as a source
of asymmetry for the density of the lepton doublets $Y_{\lepton_i}$, related to $Y_{L_i}$ by $Y_{L_i} = Y_{\lepton_i}
+ Y_{e_i}$, where $Y_{e_i}$ is the asymmetry of the right-handed charged leptons. If we only consider the
$\lepton_i$-violating interactions, the Boltzmann equations for the $Y_{\lepton_i}$ have the (linearized) form
\be
\label{eveqs}
\dot{Y}_{\lepton_i} = S_i - \gamma_i Y_{\lepton_i},
\ee 
where $\dot{Y}_{\lepton_i} \equiv {\rm d} Y_{\lepton_i}/{\rm d}z $, $z \equiv T/\MN $, and both the sources $S_i$
and the wash-out coefficients $\gamma_i$ are functions of $z$, properly normalized to account for the universe
expansion.

The way the evolution equations~(\ref{eveqs}) are converted into evolution equations for the asymmetries  
$Y_{\Delta_i}$~\cite{olive}, defined in~(\ref{Deltai}), depends on which interactions are fast at $T^*$, the decay temperature of $N$.
Three temperature intervals are of interest:
\begin{itemize}
\item[i)] $T \circa{>} \Tsph \GeV$, where the gauge interactions are in equilibrium
while no lepton Yukawa coupling mediates equilibrium interactions.
\item[ii)] $10^9 \GeV \circa{<} T \circa{<} \Tsph \GeV$, where gauge interactions,
the interactions due to the $t,b,c,\tau$-Yukawa couplings
and (presumably) the sphaleron interactions are in equilibrium~\cite{sphaleronOld}\footnote{
The rate of sphaleron interactions at finite temperature is uncertain.
More recent works~\cite{sphaleronNew} indicate an equilibrium temperature close to $10^{10}\GeV$, lower than the
value adopted here.
If this is confirmed, it is straightforward to modify the following considerations.}.

\item[iii)] $T \circa{<} 10^9 \GeV$, where also the $\mu$ and $s$-Yukawa couplings
and the CKM mixings mediate equilibrium rates.
\end{itemize}
The electron Yukawa interactions can be neglected as they are in equilibrium only at very low temperatures ($T \circa{<} 30 \TeV$)~\cite{olive}.
If we neglect the $Y_{\lepton_i}$-violating interactions in eq.~(\ref{eveqs}),
the different temperature intervals are characterized by different conservation laws,
$\dot{Y}_{Q_\alpha}=0$. At the highest temperature (case i) the (globally) conserved quantities, other than 
$Y_{\Delta_i}$, include $Y_B$, $Y_{\lepton_i}$ and $Y_{e_i}$, which are progressively broken at lower temperatures. 
The different fast rates in any temperature interval give rise to different equilibrium conditions for the asymmetries
in the number densities, or the chemical potentials $\mu_p$~\cite{mucalcoli,olive}.
For example
$\mu_{\ell_\tau}-\mu_{e_\tau}+\mu_{\higgs}=0$
in the temperature range where the $\tau$ Yukawa coupling $\lambda_\tau {\ell}_\tau \bar{e}_\tau \higgs$ mediates a fast rate.
These equilibrium conditions allow to express all the 
asymmetries in the number densities in terms of the $Y_{Q_\alpha}$, with  expressions which depend on the temperature interval.

It is now easy to see how the evolution equations~(\ref{eveqs}) in presence of lepton number violation
are converted into evolution equations for the $Y_{\Delta_i}$, directly relevant to the evaluation
of the baryon asymmetry through~(\ref{conversion}). 
Since the fast interactions conserve $\Delta_i$, they do not contribute to the equation for $Y_{\Delta_i}$,
which takes the form
\be
\dot{Y}_{\Delta_i} = - ( S_i - \gamma_i Y_{\lepton_i}),
\ee
where $Y_{\lepton_i}$ have to be expressed in terms of the $Y_{\Delta_i}$ themselves, $Y_{\lepton_i} = A_{ij}(T)Y_{\Delta_j}$.
Using the equilibrium conditions relevant to the different temperature intervals,
the matrices of constant coefficients are given by
\begin{eqnsystem}{sys:matrici}
\hbox{i)}\makebox[7cm][r]{ $A(T \circa{>} \Tsph \GeV)$}  &=&  - \diag(1,1,1); \\
\label{tauineq}
\hbox{ii)}\makebox[7cm][r]{$A(10^9 \GeV \circa{<} T \circa{<} \Tsph \GeV)$} & =&  
\pmatrix{-317/351 & 34/351 & 20/351\cr 34/351 & -317/351 & 20/351 \cr 1/117 & 1/117 & -82/117}
\label{muetauineq}; \\
\hbox{iii)}\makebox[7cm][r]{$A(T \circa{<} 10^9 \GeV)$} & =&\makebox[8cm][l]{$ 
\pmatrix{-218/253 & 25/253 & 25/253\cr 29/506 & -493/759 & 13/759 \cr 29/506 & 13/759 & -493/759}
$.}\end{eqnsystem}
If the decay of the heavy neutrino occurs at the border of any of these temperature intervals,
a more accurate treatment of the evolution equations would be needed.

\mysection{Flavour properties of leptogenesis by heavy neutrino decay}\label{flavour}
The model of neutrino masses is specified by the Yukawa couplings of the heavy right-handed neutrinos $N_\alpha$,
$\alpha = 1,2,3$, to the left-handed SU(2)-doublets $\lepton_i$, $i=e,\mu,\tau$, and to the Higgs field $\higgs$ 
\be
\label{Yukawa}
{\cal{L}}_Y = \lambda_{i \alpha} \bar \lepton_i N_\alpha \higgs + {\rm h.c.}
\ee 
which one can choose to write in the physical flavour basis of both the charged leptons and the heavy neutrinos of
mass $M_\alpha$. In this flavour basis, the mass matrix of the light neutrinos is ($v \equiv \md{\higgs}=174\GeV$)
\be
\label{lightm}
m_{ij} = \lambda_{i \alpha}^* \frac{v^2}{M_\alpha} \lambda_{j \alpha}^* . 
\ee
When needed, we accept here the currently dominant view that the light neutrino masses are hierarchical and that 
the two signals interpretable as due to the oscillations of three neutrinos are those related to the 
atmospheric and solar anomalies. In this context, the likely ranges for the eigenvalues of~(\ref{lightm}), 
$m_{\nu_3} > m_{\nu_2} > m_{\nu_1}$, are
\begin{equation}
\label{matm} m_{\nu_3} = (\Delta m^2_{\rm atm})^{1/2} \approx (0.03 \div 0.1) \eV;\qquad
m_{\nu_2} = (\Delta m^2_{\rm sun})^{1/2} \circa{<} 10^{-2} \eV
\end{equation}
The flavour structure of~(\ref{Yukawa}) and~(\ref{lightm}) is such that also the eq.s~(\ref{eveqs}) 
for $\dot{Y}_{\lepton_i}$ are different in the various ranges of $T^*$ ($\approx \MN $).
As in the connection between $Y_{\lepton_i}$ and $Y_{\Delta_i}$, what counts are the temperature ranges described in the previous section: 
we discuss them in turn. For the time being, we consider the asymmetry produced by the decay of the lightest right-handed 
neutrino $N_1$.

\subsection{$T^* \approx \MN  \circa{>} \Tsph \GeV$}  
In this case the Yukawa and the sphaleron interactions can be approximately neglected during $N_1$-decay.
Disregarding first the $\Delta L = 2$ interactions, the evolution equation for $Y_{\lepton}$ has the usual form
discussed in the literature 
\be
\label{nodelta2}
\dot{Y}_{\lepton} =S-\gamma Y_{\lepton},
\ee
where $S$ is proportional to the total decay asymmetry of $N_1$
\be
\label{asymmetry}\label{eq:epsN}
\epsN = \frac{\Gamma(N_1 \rightarrow \lepton \bar\higgs) - \Gamma(N_1 \rightarrow \bar\lepton \higgs)}
{\Gamma(N_1 \rightarrow \lepton \bar\higgs) + \Gamma(N_1 \rightarrow \bar\lepton \higgs)} =
\frac{1}{8 \pi}
\frac{1}{(\lambda^\dagger \lambda)_{11}} \sum_{\alpha = 2,3} {\rm Im} \left[(\lambda^\dagger \lambda)^2_{1\alpha}\right]
f\left(\frac{M_\alpha}{\MN }\right)
\ee
and $f(x) \simeq - 3/2x$ for $x \gg 1$~\cite{epsN}.
As long as the Yukawa interactions are negligible the lepton 
asymmetry can be considered concentrated on the $\lepton_1$ state,
the lepton doublet to which $N_1$ couples in the tree-level approximation:
\be
\label{N1coupled}
\lepton_1 \equiv  \lambda^*_{i1} \lepton_i/\sqrt{(\lambda^\dagger \lambda)_{11}}.
\ee
In an arbitrary flavour basis, the lepton asymmetry is described by a $3\times3$ matrix $\rho$ arising from the
difference of the density matrices of leptons and anti-leptons and normalized so that $\Tr\rho=\sum_i Y_{\lepton_i}$~\cite{rho}.
In the case we are considering, as shown below, $\rho$ is proportional to the projector $P_1$ on the state $\lepton_1$ up to non-diagonal
contributions.
When the Yukawa interactions come into equilibrium at lower temperatures, they
simply kill the off-diagonal terms of the matrix $\rho$ in the physical lepton-flavour basis, while leaving 
unaltered the trace, which is what influences $\sum_i Y_{\Delta_i}$ or, ultimately, the baryon asymmetry.

Let us now come to discuss the possible washing effect of the $\Delta L = 2$ interactions mediated by 
off-shell $N_\alpha$-exchanges. Their rates, at $T < \MN $, are controlled by the light neutrino mass matrix~(\ref{lightm}).
Focusing on the heaviest eigenvalue, (\ref{matm}), the dominant $\Delta L = 2$ amplitude will act on the state
$\nu_3$ and will be proportional to $m_{\nu_3}/v^2$. The corresponding rate is in equilibrium at 
temperatures above $\Tsph \GeV$, so that it is only in this range that the lepton asymmetry produced by $N_1$
is affected. This effect is readily included in the evolution equation for $Y_{\lepton}$, eq.~(\ref{nodelta2}), if 
$\lepton_3$ coincides with $\lepton_1$, eq.~(\ref{N1coupled}), as it happens if the exchange of $N_1$ dominates the mass
matrix~(\ref{lightm}) of the light neutrinos. This is usually done in the literature~\cite{L-SM} by introducing 
a properly normalized $\Delta L =2$ rate $\gamma_{\Delta L = 2}$ in the left-hand-side of eq.~(\ref{nodelta2}),
which becomes
\be
\label{eq:condelta2}
\dot{Y}_{\lepton} =S- (\gamma + \gamma_{\Delta L = 2}) Y_{\lepton}.
\ee
The inclusion of the $\Delta L = 2$ washing effect is different if $\lepton_3$ is not aligned with $\lepton_1$.
If we keep only the dominant $\Delta L = 2$ washing interactions acting on $\lepton_3$,
their effect is accounted for, in the general case, by projecting  the evolution equation for the matrix $\rho$
in the subspace $\lepton_3$ and in its orthogonal complement, and writing $\sum Y_{\lepton_i}=Y_3+Y_\perp$ with
\begin{eqnsystem}{sys:T>12}
\dot{Y}_3 & = & \ratio{3} S- [\Tr(P_1P_3)\gamma + \gamma_{\Delta L = 2}] Y_3 \label{spazio3}  \\
\dot{Y}_\perp & = & \left( 1 -\ratio{3}  \right)S- \gamma[1-\Tr(P_1P_3)]Y_\perp. \label{spazioort}
\end{eqnsystem}
Here $Y_\perp$ describes the trace of the matrix $\rho$ restricted to the corresponding subspace,
\be\label{eq:c3}
\ratio{3} = \frac
{\Gamma(N_1 \rightarrow \lepton_3 \bar\higgs) - \Gamma(N_1 \rightarrow \bar\lepton_3 \higgs)}
{\Gamma(N_1 \rightarrow \lepton   \bar\higgs) - \Gamma(N_1 \rightarrow \bar\lepton   \higgs)}
\ee 
is the part of the asymmetry produced by the decay of $N_1$ into $\lepton_3$ and $P_1$, $P_3$ are the projectors
over $\ell_1$, $\ell_3$ respectively.

This intuitive result can be formally derived by writing the evolution equation for the matrix $\rho$ associated with the lepton 
asymmetry
\be
\label{eqcondeltaP}
\dot{\rho} =S\big(P_1 + \frac{\Delta P}{2 \epsN}\big) - \gamma \frac{\{
P_1 , \rho\}}{2}  - \gamma_{\Delta L = 2} \frac{\{P_3 , \rho\}}{2},
\ee
where 
$\Delta P / 2 \epsN$ in the source term accounts for the misalignment in flavour space of the states
$\lepton$ and $\bar \lepton$ to which $N_1$ decays, with projectors $P$ and $\bar P$ (see appendix~\ref{Pappendix}).
An exact expression for the source term is
\be
S\:\frac
{\Gamma(N \rightarrow \lepton \bar\higgs) P - \Gamma(N \rightarrow \bar\lepton \higgs) \bar P}
{\Gamma(N \rightarrow \lepton \bar\higgs)   - \Gamma(N \rightarrow \bar\lepton \higgs) }, 
\ee 
which, expanded to first order in the asymmetry, gives the source term in~(\ref{eqcondeltaP}) with 
$\Delta P = P- \bar P$ and $\epsN$ given by eq.~(\ref{asymmetry}). It is important to notice that
$\Delta P / 2 \epsN$
is of order unity and cannot be neglected. Without $\gamma_{\Delta L =2}$,
since $\frac{1}{2} \{ P_1 , \Delta P\} = \Delta P + {\cal{O}}(\epsN^2)$, the solution of~(\ref{eqcondeltaP})
to first order in the asymmetry is
\be
\label{rho}
\rho = Y_{\lepton} \left(P_1 + \frac{\Delta P}{2 \epsN}\right)
\ee 
where $Y_{\lepton}$ satisfies~(\ref{nodelta2}). Notice that $\Tr \rho = Y_{\lepton}$ and that
$\Tr[ P_1 (P_1 + {\Delta P}/{2 \epsN}) ] = 1 + {\cal{O}}(\epsN)$. 
With $\gamma_{\Delta L =2}$, the restriction of eq.~(\ref{eqcondeltaP}) to the two orthogonal spaces done before leads 
to eq.s~(\ref{spazio3}) and~(\ref{spazioort}) since (see appendix~\ref{Pappendix}) 
\be
\label{restriction}
\Tr\left[P_3 \left(P_1 + \frac{\Delta P}{2 \epsN}\right)\right] = \ratio{3}.
\ee

\subsection{$10^9 \GeV \circa{<}\: T^* \approx \MN  \:\circa{<} \Tsph \GeV$}
In this case, during $N_1$-decay, the $\Delta L = 2$ interactions can be neglected.
On the contrary, one has to take into account the fast rates due to the $\tau$-Yukawa interactions. 
One of their
effects is best seen in the flavour basis, where they drive to zero the off-diagonal terms $\rho_{\tau e}$,
$\rho_{\tau \mu}$ of the matrix $\rho$. 
Defining $\ratio{\tau}$ in analogy with $\ratio{3}$, eq.\eq{c3},
one would have
\be
\dot{\rho}_{\tau \tau} = \dot{Y}_{\lepton_\tau} =\ratio{\tau} S - \gamma
\Tr (P_1 P_\tau) Y_{\lepton_\tau}
\ee  
and, for the sum of the electron and muon asymmetries, indistinguishable at this stage,
\be
\dot{\rho}_{\mu \mu} + \dot{\rho}_{e e} = \dot{Y}_{\lepton_\mu} + \dot{Y}_{\lepton_e} = 
( 1 - \ratio{\tau}) S  - \gamma [1- \Tr (P_1 P_\tau)] (Y_{\lepton_\mu} + Y_{\lepton_e}).
\ee
Furthermore, the equilibrium of the $\tau$-Yukawa and sphaleron interactions has the effect
described in section~\ref{setting} and accounted for by the $2\times 2$ matrix $A_{nm}$ which mixes $\Delta_\tau$
with $\Delta_e + \Delta_\mu$ ($n,m=\{e+\mu,\tau\}$). In analogy with section~\ref{setting}
\be
A_{nm}=
\pmatrix{(34-317)/351 & 40/351\cr 1/117 & -82/117}.
\ee

\subsection{$T^* \approx \MN  \circa{<} 10^9 \GeV$}
The extension to this case, when both the $\tau$-Yukawa and $\mu$-Yukawa interactions are in equilibrium, is obvious.
All the off-diagonal elements of the matrix $\rho$ in the flavour basis are driven to zero so that
\be
\dot{Y}_{\lepton_i} =\ratio{i}S - \gamma {\rm Tr}(P_1 P_i) Y_{\lepton_i}
\ee 
where $i = e, \mu, \tau$.
Including the fast Yukawa and sphaleron interactions as described in section~\ref{setting}
with the proper ``mixing'' matrix~(\ref{muetauineq})
one obtains the evolution equations for $Y_{\Delta_i}$.

\setlength{\unitlength}{1cm}
\begin{figure}[t]
\begin{center}
$$\hspace{-5mm}\includegraphics[width=8cm]{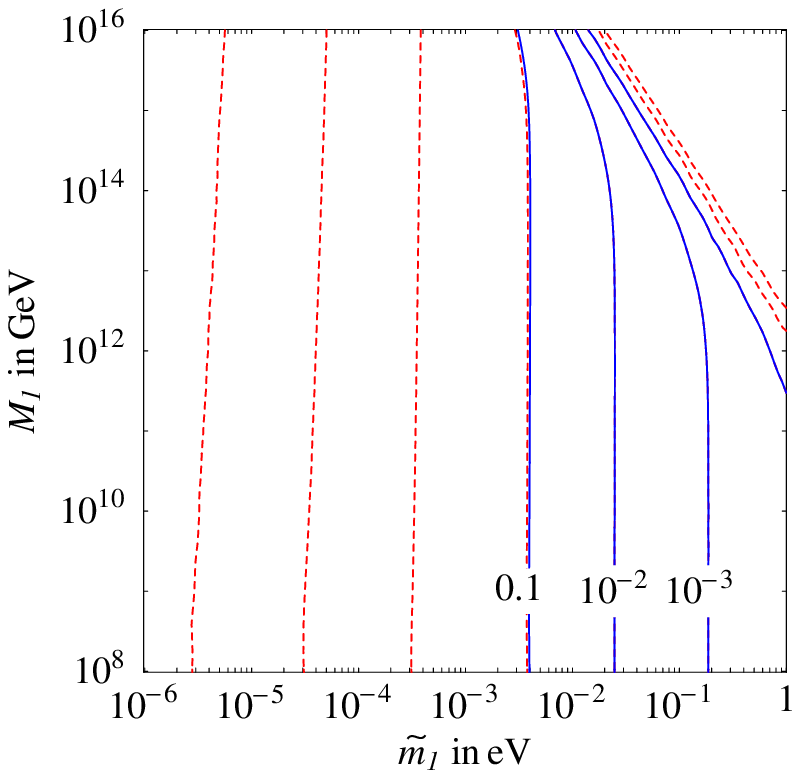}\hspace{9mm}
{\includegraphics[width=8cm]{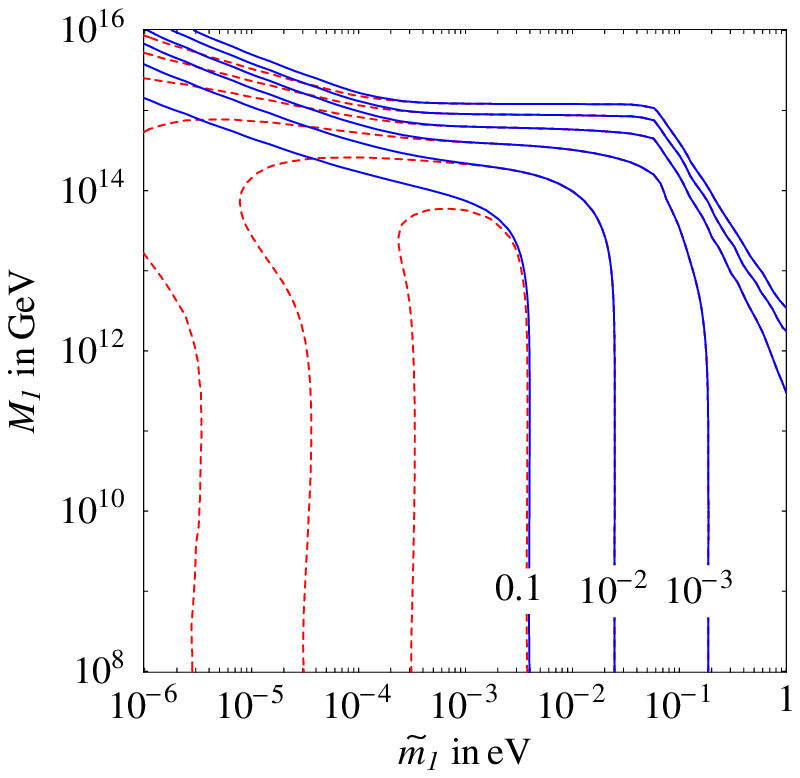}}$$
\begin{picture}(16,0)
\put(3.5,9){fig.\fig{Ex}a}
\put(12.5,9){fig.\fig{Ex}b}
\put(10.5,3.5){A}
\put(15.3,3.5){B}
\put(15.3,8){C}
\put(10.5,8){D}
\end{picture}
\vspace{-8mm}
\caption[SP]{\em Contour plot of the efficiency factor $\eta$ of leptogenesis
in the SM as function of $(\mNN,\MN $).
The $\Delta L=2$ interactions not mediated  by $N_1$ have been computed assuming $m_{\nu_3}=
\max(\mNN,(3r~10^{-3})^{1/2}\eV)$
and assuming that their flavour structure gives
the weakest washing ($X=1$) in fig.~\ref{fig:Ex}a,
and the strongest washing ($X=m_{\nu_3}^2/\mNN^2$) in fig.~\ref{fig:Ex}b.
Continuous (dashed) lines assume thermal (zero) initial abundance of $N_1$.
\label{fig:Ex}}
\end{center}\end{figure}

\mysection{Approximate analytic solutions of the Boltzmann equations}\label{approx}
In this section, and in the related appendix~\ref{mostro}, we describe an approximate but sufficiently accurate analytic solution to
Boltzmann equations of the form\eq{condelta2}.
Numerical solutions have been presented in~\cite{L-SM},
where the relevant SM rates have been computed.
As previously discussed, this equation is appropriate for $T^*>\Tsph\GeV$ and when the exchange of $N_1$ dominates the mass
matrix of the light neutrinos.
Boltzmann equations with a similar structure hold at lower temperatures, when sphaleron and Yukawa interactions are fast.
Similar analytic solutions can be developed in these cases.

In the most general case, $N_1$ exchange gives a contribution
$\tilde{m}_1 [(\lepton_1 h^*/v)^2+\hbox{h.c.}]$ to the light neutrino mass operator\footnote{Our parameter
$\tilde{m}_1$ is renormalized at the high energy scale $\sim \MN$.
High energy neutrino masses are $r\approx 1.2\div 1.3$ times larger than at low energy~\cite{SMRGE,thermal}.}:
the flavour of $\ell_1$ can be different
from the flavour of the heaviest neutrino $\nu_3$ in the $\ell_3$ doublet,
and $\mNN=(\lambda^\dagger \lambda)_{11}v^2/\MN $ can be smaller than its mass $m_{\nu_3}$.
The appropriate system of eq.s\sys{T>12} depends on the relative flavour between $\ell_1$ and $\ell_3$.
They simplify to a single equation of the form\eq{condelta2} in the two extreme limits:
\begin{itemize}

\item[(a)] $\ell_1$ and $\ell_3$ have `orthogonal' flavours (in the notation of section~\ref{flavour} this means $\Tr(P_1P_3)=0$ and $\ratio{3}=0$);

\item[(b)] $\ell_1$ and $\ell_3$ have `aligned' flavours ($P_1=P_3$ and $c_3=1$).

\end{itemize}
In case (a) the dominant $\Delta L=2$ interactions cannot wash out the leptonic asymmetry in $\ell_1$,
as would happen for example if they only acted on $\mu$ and $\tau$ flavours while $\ell_1=\ell_e$.
The relevant equation~(\ref{spazioort}) has the form\eq{condelta2}, with $\gamma_{\Delta L=2}$ given by the
subdominant $\Delta L=2$ interactions mediated by $N_1$ (neglected in\sys{T>12}).
In case (b) the relevant equation~(\ref{spazio3}) has the form\eq{condelta2},
with the $\Delta L=2$ rate $\gamma_{\Delta L=2}$ being $X=(m_{\nu_3}/\mNN)^2$ times stronger
than the one mediated by $N_1$ alone.
The intermediate situation can be studied by solving\sys{T>12} for any given
relative flavour misalignement between $\ell_1$ and $\ell_3$:
the result should be intermediate between the ones obtained for $X=1$ (case a) and $X=m_{\nu_3}^2/\mNN^2$ (case b).




Both the source $S$ and the decay coefficient $\gamma+\gamma_{\Delta L=2}$ in eq.\eq{condelta2} depend upon $Y_{N_1}$, 
the $N_1$ density relative to the total entropy density, which satisfies its own evolution equation.
We define (see appendix~\ref{mostro})
\be\label{eq:mstar}
\mNNstar\equiv 2.3~10^{-3}\eV.
\ee 
If $\mNN\ll\mNNstar$, the decay width of $N_1$ is much smaller than the Hubble constant and $N_1$ decays strongly out of equilibrium,
while, conversely, it is almost in equilibrium for $\mNN\gg\mNNstar$. Explicit expressions for $Y_{N_1}(z)$ in the two regimes 
are given in appendix~\ref{mostro}.

With explicit knowledge of $S$ and $\gamma$, eq.\eq{condelta2} can be integrated to obtain $Y_{\lepton}$ at $T \ll \MN $
\begin{equation}\label{eq:2183}
Y_{\ell}(z\sim\infty)=\epsN Y_{N_1}^{\rm eq}(0)\eta\qquad\hbox{i.e.}\qquad
\frac{n_B}{s} = -\frac{12}{37} Y_{\ell}(z\sim\infty) = -\frac{3\epsN \eta}{2183},
\end{equation}
where $\eta \leq 1$ is an efficiency factor describing the effect of wash-out interactions.
In appendix~\ref{mostro} we describe an accurate analytic approximation for $\eta$.
The first-order linear differential equation for $Y_{\ell}$, eq.\eq{condelta2},
can be explicitly solved in terms of integrals.
Our approximation for $\eta$ is obtained by inserting the asymptotic expressions
for the interaction rates and for the $N_1$ abundance described in appendix~\ref{BN1}
and evaluating the integral in the saddle-point approximation.

Depending on the mass and Yukawa couplings of $N_1$, four different regimes are possible.
The $\Delta L=2$ interactions can be either significant or negligible, and the $N_1$ decay rate
(which is comparable to the thermally averaged $\Delta L=1$ interactions)
can be faster or slower than the expansion of the universe, depending on the value of $\mNN/\mNNstar$.
Specializing our general approximation\eq{PuntoSella} described in the appendix to the four regimes, we can derive less precise but 
more explicit expressions for $\eta$:
\begin{itemize}

\item[A.] $N_1$ decays strongly out of equilibrium and all the wash-out interactions are negligible.
In this case \Blue$\eta \approx 1$\Black.

\item[B.] $N_1$ decays almost in equilibrium and $\Delta L=1$ interactions are not negligible.
As explained in appendix~\ref{mostro}, $\eta$ is suppressed only almost linearly in $\mNNstar/\mNN$:
\begin{equation}
\Blue\eta \approx 0.168\frac{\mNNstar/\mNN}{\sqrt{\ln(\mNN/\mNNstar)}}\Black
\end{equation}
We put a numerical coefficient  $10\%$ lower than what suggested by\eq{PuntoSella}
so that this approximation agrees with our full computation within few $\%$.
Notice that $\eta$ depends, up to small corrections, on the mass and Yukawa couplings of $N_1$ only through the single combination 
$\mNN$.

\item[C.] $N_1$ decays almost in equilibrium and $\Delta L=2$ interactions are not negligible.
These interactions wash out $Y_{\ell}$ exponentially:
\begin{equation}
\Blue \eta \approx \exp\bigg[-\frac{\mNN}{\mNNstar}\sqrt{\frac{M_1X}{3.3~10^{15}\GeV}}\bigg].\Black
\end{equation}
The factor in the exponent is proportional to the square root of $\gamma_{\Delta L=2}$ and gives
a significant suppression  only if $M_1$ is heavy enough.

\item[D.] $N_1$ decays strongly out of equilibrium and $\Delta L=2$ washing is not negligible.
These interactions again wash out $Y_{\ell}$ exponentially:
\begin{equation}
\Blue \eta \approx \exp\bigg[-\bigg(\frac{\mNN}{\mNNstar}\bigg)^{1/3}\bigg(\frac{M_1}{0.8~10^{13} \GeV}
\frac{X}{m_{\rm \nu_3}^2/\mNN^2 }\bigg)^{2/3} \bigg].\Black
\end{equation}
If $X=1$, this suppression is significant only for uninteresting high values of $M_1$.

\end{itemize}
These four regions are clearly visible, and explicitly indicated, in fig.~\ref{fig:Ex}, where we show contour plots of 
 $\eta$, as function of
the two unknown parameters $\mNN$ and $\MN $ in the two extreme situations alluded to above:
fig.\fig{Ex}a refers to case (a) and
fig.\fig{Ex}b to case (b).

Before concluding this section, we make a few comments on the validity of our approximations.
\begin{enumerate}
\item We have assumed that $N_1$ has a thermal abundance at $T\gg \MN $ but this could not be the case.
If $\mNN\gg \mNNstar$, $N_1$ rapidly thermalizes anyhow, so that its `initial' abundance is irrelevant.
If instead $\mNN\ll \mNNstar$, thermalization is slow and leptogenesis becomes sensitive to the initial abundance:
starting with $Y_{N_1}(0) = 0$, the $N_1$ abundance at decay is roughly suppressed by a factor $\mNN/\mNNstar$,
not included in fig.s~\fig{Ex}.

\item In presence of a strong exponential suppression, the slow (i.e.\ $\gamma\ll 1$) Yukawa and/or sphaleron interactions
can not be neglected:  a fraction $\eta \sim \gamma/\gamma_{\ell}$ of the asymmetry in left-handed leptons $\ell$
is transmitted to right-handed leptons and/or to quarks before being washed out.

\item We have done the computation in the SM.
Apart from factors of order one, there should be no significant difference between the
SM~\cite{L-SM} and MSSM~\cite{L-MSSM} predictions for $Y_\ell$,
with the main change possibly due to a large value of  $\tan\beta$.

\item We have neglected the asymmetry generated by the heavier neutrinos, even if their mass is below $T_{\rm rh}$.
This asymmetry is generally believed to be washed out by fast interactions mediated by the lightest right-handed neutrino.
Considering the flavour structure of the problem we can see
that this is not always the case. The heavier neutrinos produce an asymmetry of the generic form (see eq.~(\ref{rho}))
\be
\rho_h = Y_h \left(P_h + \frac{\Delta P_h}{2 \varepsilon_h}\right).
\ee
For $T^* \circa{>} \Tsph \GeV$, analogously to eq.~(\ref{spazio3}) and~(\ref{spazioort}), only the restriction 
of $\rho_h$ to the subspace $\lepton_1$ will be washed out by $N_1$ interactions, while its orthogonal complement will 
remain untouched. For lower $T^*$ the wash-out efficiency depends on the projection of $\lepton_h$ and $\lepton_1$ on the 
flavour states: also in this case a non-negligible part of the asymmetry can survive. 

\end{enumerate}

\mysection{Baryogenesis in specific models of neutrino masses}
As made clear from the previous discussion, a detailed calculation of the baryon asymmetry generated via 
leptogenesis would require a detailed knowledge of the model for neutrino masses, not available at present. A typical model
is currently specified by textures for the matrix $\lambda$ of the Yukawa couplings and for the mass matrix $M$ of the
heavy right-handed neutrinos, with undetermined complex coefficients of order unity for every non-zero entry. The 
ignorance of these coefficients induces an uncertainty in the lepton asymmetry, generically of the same order unity. 
A pile up of effects correcting the final result for $Y_\lepton$ by up to one order of magnitude in either direction,
although not likely, cannot be safely excluded.

\lascia{
Before considering specific models, we study the impact of experimental data on the
parameters that affect leptogenesis: the masses and Yukawa couplings of heavy right handed neutrinos $\NR$
(see~\cite{altri studi} for related discussions).
The lightest $\NR$ whose decays generate the leptonic asymmetry
gives a contribution $\mNN$  to the light neutrino mass matrix that
may or may not be the dominant one, $m_{\nu_3}$.
\begin{enumerate}
\item If $\mNN\ll m_{\nu_3}$ the efficiency of leptogenesis is close to maximal, $\eta\approx 1$ (unless $\MN $ is very heavy or
the $N_1$ abundance is much smaller than the thermal one).
Since a heavier $\NR$ must mediate the experimentally measured $m_{\nu_3}$, the CP asymmetry in $N_1$ decays
can be estimated as
$\epsN\approx -\MN /10^{16}\GeV$ (eventually times a flavour factor less than one).
Consequently, the experimentally observed baryon asymmetry is obtained for
an appropriate value of $\MN $: $(n_B/s)/10^{-10}=\MN /10^{9}\GeV$.

\item If instead $\mNN=m_{\nu_3}$ the efficiency of leptogenesis is $\eta\sim 0.01$ (or smaller if $\MN \circa{>}10^{14}\GeV$).
The heavier $\NR$ give a contribution to the neutrino mass matrix $r$ times smaller
than the dominant one, with $r$ related to the still unknown solar mass splitting.
Consequently $\epsN\approx -\MN  r/10^{16}\GeV$ (eventually times a flavour factor less than one)
and $(n_B/s)/10^{-10}\approx r \MN /10^{11}\GeV$: 
unless $r$ is too small the observed baryon asymmetry can be obtained for $\MN \approx 10^{11\div 14}\GeV$.
\end{enumerate}}

Before considering specific models, it is useful to recast the relevant equations in a more expressive form,
emphasizing the role of the relevant parameters~\cite{altri studi}.
From\eq{epsN}, the total decay asymmetry can be written as
\begin{equation}\label{eq:g}
|\epsN|=\frac{3}{16\pi}\frac{m_{\nu_3}\MN }{v^2}g
\end{equation}
where $g$ is a model dependent flavour factor which, barring cancellations, is less than unity
and does not depend on any absolute scale.
Ref.~\cite{DI} shows that $g\le1$ if  $\MN  \ll M_{{2,3}}$.
We find therefore, from\eq{2183}
\begin{equation}\label{eq:YB/YBexp}
\frac{n_B/s}{10^{-10}}\approx \eta g \frac{\MN }{10^9\GeV},
\end{equation}
where the efficiency factor $\eta$ is also less than unity and depends on the coefficient $\mNN$ of the
contribution
$\tilde{m}_1 [(\lepton_1 h^*/v)^2+\hbox{h.c.}]$ to the mass matrix of the light neutrinos from $N_1$ exchange.

In the light of the above general considerations,
we compute the expected baryon asymmetry in two models of neutrino masses, based on an abelian U(1) or
a non-abelian U(2) family symmetry respectively. In both cases we have checked that the careful 
treatment of flavour, as described in section~\ref{setting} and~\ref{flavour}, can modify the result with respect
to the naive treatment based on eq.~(\ref{eveqs}) only to an amount which stays within the model uncertainty.
Hence we only describe the results of the naive analysis. As in all the previous discussion we assume a reheating 
temperature after inflation above $\MN $.

\subsection{The U(1) model}
The essential elements of this model~\cite{U1} are the U(1)-charges of the three left-handed lepton doublets 
$\tilde\lepton_3$, $\tilde\lepton_2$,
$\tilde\lepton_1$: $a$, $a$, $a+x$, and the biggest charge $\theta_1$ of the heavy neutrinos,
associated with $\tilde{N}^c_1$.
The tilde on top of $\ell$ and $N^c$ is there to remind that they are not mass eigenstates.
The U(1) charges of all particles with the same chirality --- hence the `$^c$' on the heavy neutrino fields ---
are non-negative and are taken consistent with SU(5)
unification, so that they are identical within each SU(5) multiplet. Every element $\lambda_{i \alpha}$ of the Yukawa matrix
is given, up to a factor of order unity, by $\delta^{q_i + \theta_\alpha}$, where $\delta$ is a small parameter and $q_i$,
$\theta_\alpha$ are the U(1) charges of the fields $\tilde\lepton_i$ and $\tilde{N}_\alpha^c$ respectively. An analogous formula holds for all 
other Yukawa couplings. In the same way, again up to factors of order unity, the mass matrix of the heavy neutrinos
is $M_{\alpha \beta} \approx M_0\:\delta^{(\theta_\alpha + \theta_\beta)}$, where $M_0$ is an unspecified mass scale. The equality
of the U(1)-charges of $\tilde\lepton_3$ and $\tilde\lepton_2$ is at the origin of the large neutrino mixing supposedly observed at
SuperKamiokande. The large top Yukawa coupling and consistency with SU(5) unification require the vanishing of the U(1)
charge of the third generation right-handed charged leptons, approximately $\tau_R$.

With these inputs, from the explicit construction of the $\lambda$ and $M$ matrices, it is immediate to obtain
\be
\label{U1data}
\epsN \approx \frac{3}{16 \pi} \delta^{2(a+\theta_1)},\quad m_{\nu_3} \approx \widetilde{m}_1 \approx 
\delta^{2a} \frac{v^2}{M_0},\quad \MN  \approx \delta^{2 \theta_1} M_0.
\ee 
Phases of order unity are assumed in the various matrix elements. We have also assumed that the unspecified charges 
of the heavy neutrinos other than $N_1$ are sufficiently smaller than $\theta_1$ so that $\MN  / M_\alpha \approx \delta^
{2(\theta_1-\theta_\alpha)}$, $\alpha = 2,3$, are small.
A posteriori, this can only be marginally the case.
A drawback of the model is the need to invoke a fortuitous cancellation
to explain the hierarchy in the neutrino mass matrix, $m_{\nu_2}/m_{\nu_3} \circa{<} 0.05$, against the naive expectation
$m_{\nu_2} \approx m_{\nu_3}$. In its unified version, the model also gives $m_\tau \approx \delta^a v$, although with
an arbitrary identification of the vacuum expectation values of the Higgs fields relevant to $m_\tau$ and $m_{\nu_3}$.

Eq.s~(\ref{U1data}) can be combined to give the flavour factor $g$ in\eq{g}, approximately equal to unity.
Furthermore $\MN $, using also $m_\tau \approx \delta^a v$, is bounded by $\MN  < M_0 \approx m_\tau^2 / m_{\nu_3} < 10^{11} \GeV$.
There is no other restriction on $\MN $ which
can therefore be fixed to obtain the desired value of $Y_B$ from fig.~(\ref{fig:Ex}).
Since $\mNN\approx m_{\nu_3}\approx (0.3\div 1)10^{-1}\eV$, from fig.s\fig{Ex},
$\eta=(0.3\div 1) 10^{-2}$.
Hence consistency with $Y_B = (10^{-10} \div 10^{-11})$ is obtained for $\MN  \approx 10^{10\div11} \GeV$.
Although lower than $M_0\circa{<}10^{11}\GeV$, this value is such that also the heaviest singlet neutrinos
have to be close in mass and all in the $10^{11}\GeV$ range.

\subsection{The U(2) model}
The relevant pieces of information are contained in the form of the matrices for the $\tilde\lepton  \tilde{N}$ Yukawa couplings and for
the heavy neutrino masses
\be
\label{U2matrices}
\tilde\lambda \approx \left(\ba{ccc}  & \epsilon' &  \\ \epsilon' & \epsilon & \epsilon \\
 & \epsilon & 1\ea\right),\qquad
M \approx \epsilon M_0 \left(\ba{ccc} & \epsilon' & \\ \epsilon' & 1 & 1 \\  & 1 & \epsilon \ea\right),
\ee 
where $\epsilon \approx 0.02$ and $\epsilon' \approx 0.004$ are dimensionless parameters related to the 
hierarchical breaking of U(2) and determined from the observed values of charged fermion masses~\cite{U2}. The tilde on top
of $\lambda$ indicates that $\tilde\lambda$, unlike $\lambda$ in eq.~(\ref{lightm}), is not in the physical 
basis for the heavy neutrinos but rather in the basis specified by $M$. As before, order one prefactors are 
understood in every entry of~(\ref{U2matrices}), whereas the blanks stand for non-zero but sufficiently 
small elements. Since the structure of the model is intimately tied to unification, $\epsilon M_0$ is of
order $M_{\rm GUT} \approx 2 \cdot 10^{16} \GeV$~\cite{U2,U2-BHRR}.

The diagonalization of $M$ in eq.~(\ref{U2matrices}) leads to $M_2 \approx M_3 \approx \epsilon M_0 \approx M_{\rm GUT}$
and to a substantially lighter $\MN  \approx \epsilon \epsilon'^2 M_{\rm GUT} \approx 10^9 \div 10^{10} \GeV$. In turn 
the $\lepton$-$N$ Yukawa couplings in the physical N-basis ($N_1$, $N_2$, $N_3$) acquire the form 
\be
\lambda \approx \left(\ba{ccc} \epsilon'^2 & \epsilon' & \epsilon' \\ \epsilon' & \epsilon & \epsilon \\ 
\epsilon' & 1 & 1  \ea\right).
\ee  
The exchange of $\MN $ dominates the light neutrino mass matrix, giving
\be
m_{\nu_3} \approx \frac{v^2}{M_{\rm GUT}} \frac{1}{\epsilon} \approx 0.1 \eV, \qquad 
\frac{m_{\nu_2}}{m_{\nu_3}} \approx \epsilon \approx 2 \cdot 10^{-2},
\ee
in reasonable agreement with observations. Furthermore 
$\mNN\approx m_{\nu_3}$, so that $\eta\approx (0.3\div 1) 10^{-2}$ and,
for the flavour factor in\eq{YB/YBexp}, $g\approx m_{\nu_2}/m_{\nu_3}\approx \epsilon\approx 0.02$.
Putting everything together, we get from\eq{YB/YBexp}
$$\frac{n_B}{s}\approx 3~10^{-11}\frac{\MN }{10^8\GeV}.$$
Unlike the previous case, however, $\MN $ is not a free parameter.
Since $\MN \approx 10^{9\div10}\GeV$, the baryon asymmetry is fixed to $3~10^{-(12\div13)}$.

\mysection{Conclusions}
The experimental results in neutrino physics of the last few years have made plausible the case for a
baryon asymmetry generated by a lepton asymmetry arising from the out-of-equilibrium decay of a heavy right-handed neutrino.
With this motivation, this paper achieves two purposes.
The first is to determine in a precise way the evolution equations for the baryon and lepton asymmetries by taking
into account the full flavour structure of the problem.
This would be fully relevant with a sufficiently detailed model of
neutrino masses on hand.

The second purpose is to examine the expected asymmetry in two specific models
of neutrino masses, within their limit of uncertainty.
In both cases, we have shown that a consistent picture of baryogenesis can emerge,
although with significant differences.
In the U(1) model, which does not account automatically for the hierarchy of neutrino masses,
the observed baryon asymmetry can be obtained by choosing the scale of the heavy neutrinos,
all close in mass, between $10^{10}$ and $10^{12}\GeV$.
In the U(2) model, the lighter of the heavy neutrinos has a mass $\MN$ fixed by independent
considerations at $\MN \approx 10^{9\div10}\GeV$ and significantly lower than that of the two others
$M_2\approx M_3\approx \MGUT$.
In this case the baryon asymmetry is fixed at $3~10^{-(12\div13)}$ against an observed value of $10^{-(10\div 11)}$.
In view of the uncertainties discussed in the text, we consider this as a success, strengthening the view that
makes leptogenesis an appealing mechanism for baryogenesis.
In every instance it is clear that the decaying neutrino is heavy, above $10^9\GeV$
and correspondingly $T_{\rm rh}$ is bigger than the same value,
a non trivial constraint on the evolution of the early universe.

\paragraph{Acknowledgments}
The work of N.T.\ was supported by the E.C.\ under TMR contract 
No. ERBFMRX--CT96--0090
and contract No.\ ERBFMBICT983132.
A.S.\ thanks G. Giudice, A. Notari, M. Plumacher, M. Raidal and A. Riotto for useful discussions.

\appendix

\mysection{The explicit form of $N_1$ decay amplitude}\label{Pappendix}
The tree-level amplitudes and the one loop corrections to $N_1$ decay are, up to an overall factor
$$
\begin{array}{ll}
a_i\equiv  a(N_1 \rightarrow \ell_i \bar\higgs) = \lambda_{i1} \qquad&
\delta_i  \equiv  \Delta a(N_1 \rightarrow \ell_i \bar\higgs) = 
\lambda_{i \alpha}\lambda_{j 1}^* \lambda_{j \alpha} A_\alpha\\
\bar a_i  \equiv  a(N_1 \rightarrow \bar\ell_i \higgs) = \lambda^*_{i1} &
\bar\delta_i  \equiv  \Delta a(N_1 \rightarrow \bar\ell_i \higgs) = 
\lambda_{i \alpha}^*\lambda_{j 1} \lambda^*_{j \alpha} A_\alpha,
\end{array}$$
where ${\rm Im} A_\alpha = - f(M_\alpha / \MN )/ 32\pi$ and $f$ is defined in eq.~(\ref{asymmetry}). 
Note that $\bar{\delta}_i\neq \delta_i^*$.
Neglecting $\Delta L = 2$ interactions, the Boltzmann equations for the density matrices of leptons and antileptons
are
\begin{eqnsystem}{sys:app}
\label{ev1} \dot{\rho_\ell} = S_\ell P + \gamma (P - \frac{1}{2} \{P,\rho_\ell\}) \\
\label{ev2} \dot{\rho_{\bar\ell}} = S_{\bar\ell} \bar P + \gamma (\bar P - \frac{1}{2} \{\bar P,\rho_{\bar\ell}\}), 
\end{eqnsystem}
where
\be
P_{ij} = \frac{(a + \delta)_i (a + \delta)_j^\dagger}{|a+\delta|^2} \qquad
\bar{P}_{ij} = \frac{(\bar a + \bar \delta)_i (\bar a + \bar \delta)_j^\dagger}{|\bar a+\bar\delta|^2}.
\ee
By taking the difference between~(\ref{ev1}) and~(\ref{ev2}), defining $S = S_\ell - S_{\bar\ell}$
and linearizing in $\rho = \rho_\ell - \rho_{\bar \ell}$,
eq.~(\ref{eqcondeltaP}) follows.
In eq.~(\ref{eqcondeltaP}) we have also included the $\Delta L=2$ terms, which can be discussed along similar lines.
Explicitly we have
\be
\left(P_1 + \frac{\Delta P}{2 \epsN}\right)_{ij} = 
- \frac{i}{2}\frac{{\rm Im} (\lambda_{i1} \lambda_{j \alpha}^* \lambda_{k \alpha}^* \lambda_{k 1}) A_\alpha}
{\sum_i {\rm Im} (\lambda_{i1} \lambda_{i \alpha}^* \lambda_{k \alpha}^* \lambda_{k 1}) {\rm Im} A_\alpha} + {\rm h.c.},
\ee
so that eq~(\ref{restriction}) is easily justified.

\setcounter{equation}{0}
\renewcommand{\theequation}{\thesection.\arabic{equation}}

\mysection{\label{mostro}Approximate solutions of Boltzmann equations for leptogenesis}
As usual~\cite{Kolb,L-SM,L-MSSM} we assume a Boltzmann kinetic distribution for bosons and fermions:
the Boltzmann equations for the total abundances $n_{N_1}$, $n_{\ell}$ and $n_{\bar\ell}$
are expected to be correct within few $10\%$ errors.
With this approximation the Hubble constant at temperature $T$, as predicted by standard cosmology, is given by
$H^2(T)=\frac{8\pi}{3}G_N\rho$ where
$M_{\rm Pl}\equiv G_N^{-1/2}=1.22~10^{19}\GeV$,
$\rho={3g_{\rm SM}} T^4/{\pi^2}$ and
$g_{\rm SM}$ is the number of ultra-relativistic spin degrees of freedom
($g_{\rm SM}=118$ in the SM at $T\gg 100\GeV$).
The entropy density is $s=4g_{\rm SM} T^3/{\pi^2}$.
The number density of a particle $p$ with $g_p$ spin degrees of freedom
(for example $g_{N_1}=2$ and $g_{\ell_i}=2$) and mass $M_p$ in thermal equilibrium at temperature $T$ is
$$n_p^{\rm eq}(z)=\frac{g_pM_p^3}{2\pi^2}\frac{\K_2(z)}{z},\qquad
Y_p^{\rm eq}(z)\equiv {n_p^{\rm eq}\over s},\qquad
Y_p^{\rm eq}(0)=\frac{g_p}{4g_{\rm SM}}$$
where $z=M_p/T$ and $\K_2(z)$ is a Bessel function (precisely defined as {\tt BesselK[2,z]} in Mathematica notation~\cite{Mathematica})
with the following limiting behaviors:
$\K_2(z)\simeq 2/z^2$ as $z\to 0$ and
$\K_2(z)\simeq e^{-z}\sqrt{\pi/2z}$ as $z\to \infty$.
The final $B$ asymmetry can be written as
$$\frac{n_B}{s}=-\frac{12}{37} \epsN Y_{N_1}^{\rm eq}(0)\eta = -\frac{3g_{N_1}}{37g_{\rm SM}}\epsN\cdot\eta,$$
where $\epsN$ is the CP asymmetry in $N_1$ decays given in eq.~(\ref{asymmetry}), and
$\eta$ is an efficiency factor for leptogenesis that depends on the mass and Yukawa couplings of $N_1$
and should be determined solving the
relevant set of Boltzmann equations.
We assume thermal equilibrium at $T\gg \MN $: consequently $\eta\le 1$.
In this section we derive the approximate analytical expression for $\eta$ used in section~\ref{approx}.
As usual it is convenient to write equations
for $Y_{N_1}(z)\equiv n_{N_1}(z)/s(z)$ and for $Y_\ell \equiv (n_\ell -n_{\bar{\ell}})/s$,
where $z\equiv \MN /T$ and $s$ is the total entropy density.

\subsection{The Boltzmann equation for the $N_1$ abundance}\label{BN1}
We neglect corrections of relative order ${\cal O}(\lambda_t^2,g_2^2)/\pi^2$
(some of them are included in state of the art computations).
A first equation
\begin{equation}
Y'_{N_1}+\gamma_{N_1}(Y_{N_1}-Y_{N_1}^{\rm eq}) =0\qquad\hbox{where~\cite{L-SM,thermal}}\qquad
\gamma_{N_1}=\frac{\gamma_D}{szHY_{N_1}^{\rm eq}}
\end{equation}
controls  how much out of equilibrium $N_1$ decays.
All quantities depend on $z=\MN/T$.
An approximate expression for the dimensionful thermally averaged decay rate $\gamma_D$
can be found at the end of this appendix.

Assuming thermal equilibrium of $N_1$ above the decoupling temperature $T\sim \MN $,
the $N_1$ abundance at decoupling can be either almost in equilibrium (if $\gamma_{N_1}\gg 1$)
or strongly out of equilibrium (if $\gamma_{N_1}\ll 1$).
Approximately $\gamma_{N_1}\sim \mNN/\mNNstar$, where
$$\mNNstar\equiv \frac{128\sqrt{g_{\rm SM}}g_\ell v^2}{M_{\rm Pl}(1+g_{N_1})}=2.3~10^{-3}\eV$$
depends on cosmology (the numerical factors are appropriate for later use).
All the dependence on the mass and Yukawa couplings of $N_1$ is incorporated in 
$\mNN\equiv \lambda_1^2 v^2/\MN $,
the contribution to the light neutrino mass mediated by $N_1$
($\lambda_{i}^2\equiv (\lambda^\dagger \lambda)_{ii}$ is the `total' squared Yukawa coupling of $N_i$).
Approximate solutions for $Y_{N_1}$ valid in the two regimes are:
\begin{itemize}

\item If $\mNN\ll \mNNstar$, $N_1$ decays strongly out of equilibrium.
In this case the only relevant collision term is the thermally averaged decay rate at $T\ll \MN $,
so that the Boltzmann equation for $N_1$ and its solution are
$$Y'_{N_1}=-\frac{\gamma_D(z\sim\infty)}{szH}\frac{Y_{N_1}}{Y_{N_1}^{\rm eq}}\qquad\Rightarrow\qquad
Y_{N_1}(z)=Y_{N_1}^{\rm eq}(0) \exp\bigg[-\frac{\mNN}{\mNNstar}\frac{4\sqrt{2/\pi}}{3}z^2\bigg].$$
At very small $\tilde{m}_1$  $N_1$ decays slowly at $T\circa{<} m_{N_1}/g_{\rm SM}$
and gives a substantial contribution to the total energy density: 
we do not study this possibility~\cite{thermal}.

\item If $\mNN\gg \mNNstar$, $N_1$ decays almost in equilibrium.
If the collision terms $\gamma_{N_1}$ are fast,
the approximate equation for $\Delta(z)\equiv [Y_{N_1}(z)/Y_{N_1}^{\rm eq}(z)]-1$ and its solution are~\cite{Kolb}
$$\Delta'= - \gamma_{N_1} \Delta - (1+\Delta) \frac{Y^{\prime \rm eq}_{N_1}}{Y^{\rm eq}_{N_1}}
\qquad\Rightarrow\qquad
\Delta \approx -\frac{zsH}{\gamma_D} Y^{\prime \rm eq}_{N_1}.$$
with $Y^{\prime \rm eq}_{N_1} =-z^2{\rm K}_1(z)/4g_{\rm SM}$.

\end{itemize}
Even if we do not have a unique good approximation for the $N_1$ abundance,
we can obtain a good approximation for the generated lepton asymmetry.
The reason is that, if the washing interactions are effective (otherwise $\eta$ is trivially close to 1),
only the lepton asymmetry produced by late $N_1$ decays survives.
The final result is insensitive to what happens in the range of temperatures $T\approx \MN $,
where we do not have a simple approximation for $Y_{N_1}(z)$.

\subsection{The Boltzmann equation for the leptonic asymmetry}
We now consider the Boltzmann equations for the leptonic asymmetry $Y_\ell=(n_\ell-n_{\bar\ell})/s$.
For concreteness, we give approximate solutions to the equations usually employed in the literature.
If sphalerons and/or lepton Yukawa couplings and/or a non minimal flavour structure give significant interactions at $T\approx \MN $,
the equations must be modified by appropriate order one coefficients.
The Boltzmann equation for $Y_\ell$ is
\begin{equation}\label{eq:Yell}
Y'_\ell+\gamma_\ell(z) Y_\ell=S_\ell(z),
\end{equation}
where the source and damping factors $S_\ell$ and $\gamma_\ell$ are given by
\begin{equation}
S_{\ell}= \frac{\epsN\Delta \gamma_D}{szH} ,\qquad
\gamma_\ell=\frac{\gamma_D+4\gamma_N^{\rm sub}}{2szHY_{\ell}^{\rm eq}}.
\end{equation}
Approximate expressions for these factors,
inserting our almost-in-equilibrium approximation for $Y_{N_1}(z)$ (it is immediate to pass to the other case),
are:
\begin{eqnsystem}{sys:gammaApprox}
S_{\ell}(z)&\stackrel{\tilde{m}_1\gg \tilde{m}_1^*}{\approx} &- \epsN Y^{\prime \rm eq}_{N_1}
\stackrel{z\gg 1}{\approx} 
\epsN \frac{\sqrt{\pi/2}}{4g_{\rm SM}}z^{3/2}e^{-z}\\
\gamma_{\ell}(z)&\stackrel{z\gg 1}{\approx} &\frac{2\mNN}{3\mNNstar}
 \left[z^{5/2} e^{-z} +
\frac{32\sqrt{2}\ \lambda_1^2}{z^2\pi^{5/2}}X\right]\label{eq:gammaapprox}
\end{eqnsystem}
The term in\eq{gammaapprox} decoupling as $e^{-z}$ at low $T$ is due to $\Delta L=1$ interactions, while
the term multiplied by $X$ is due to $\Delta L=2$ interactions
and decouples only as $1/z^2$.
As explained in section~\ref{approx} the factor $X$, ranging between $1$ and $m_{\nu_3}^2/\mNN^2$, takes into account 
possible significant $\Delta L=2$ interactions not mediated by $N_1$.

The Boltzmann equation for $Y_\ell$ is a linear first order differential equation:
using its standard solution in terms of integrals we finally obtain
\begin{equation}\label{eq:PuntoSella}
\eta =  \int_0^\infty dz' \exp[-F(z')]\qquad\hbox{where}\qquad
F(z)\equiv -\ln\frac{S_\ell(z)}{\epsN Y_{N_1}^{\rm eq}(0)}+\int_z^\infty \gamma_{\ell}(z') dz'.
\end{equation}
Since $e^{-F}$ is a bell-shaped function,
by integrating the 2nd order Taylor expansion of $F(z)$
around its minimum ($F'(\bar{z})=0$) gives a more explicit but less
accurate expression,
$\eta \approx  e^{-F(\bar{z})}
\sqrt{{2\pi}/{F''(\bar{z}) }} $.
When $\Delta L=2$ scatterings are irrelevant this expression, with $S_\ell$ and $\gamma_\ell$ taken from\sys{gammaApprox},
gives a good approximation even if $N_1$ decays strongly out of equilibrium.

Depending on the mass and Yukawa couplings of $N_1$, four different regimes are possible.
Specializing our approximation\eq{PuntoSella} to the four regimes, we have derived the less precise but more explicit approximations for $\eta$
presented in the text.
It could be a bit surprising that, when the $\gamma_{\Delta L=1}$ interactions are much faster than the expansion rate
of the universe, the lepton asymmetry is suppressed only linearly.
The reason is that $\Delta L=1$ interactions, having $N_1$ as an external state,
at low temperature are suppressed by a Boltzmann factor:
$\gamma_{\Delta L=1}(z)\approx \gamma_{\Delta L=1}(0) e^{-z}$.
For $z\circa{>} \bar{z}$ these washing interactions become negligible and
all the $N_1$ decays give rise to unwashed leptonic asymmetry, so that the suppression factor is approximately given by
$$\eta\approx Y_{N_1}(\bar{z})/ Y_{N_1}(0) \approx e^{-\bar{z}}\approx 1/\gamma_{\Delta L=1}(0)\approx \mNNstar/\mNN.$$
The $\Delta L=2$ interactions mediated by $N_i$ can be computed in the effective theory below $N_1$ and decouple only as $1/z^2$.
These interactions do not decouple exponentially and consequently can wash out $Y_{\ell}$ exponentially.
If $\mNN\approx m_{\nu_3}$ these interactions are significant when $\lambdaN\sim 1$.

\subsection{Approximate expressions for the thermally averaged interactions rates}
Evaluating the various thermally averaged interaction rates~\cite{L-SM} is the more difficult step of a numerical computation.
Quite simple and accurate expressions hold in the low temperature limit
\begin{eqnsystem}{sys:approx}
\gamma_D(z\sim \infty)\approx \frac{\MN^4\lambdaN^2}{8\sqrt{2}\pi^{5/2}}\frac{e^{-z}}{z^{3/2}},&&\qquad
\gamma_{N}^{\rm sub}(z\sim \infty)\approx \frac{(\MN\lambdaN)^4}{\pi^5 z^6}\\
\riga{where $\lambda_i^2\equiv (\lambda\lambda^\dagger)_{ii}$
is the `total' squared Yukawa coupling of $N_i$, and
we have followed the notation of~\cite{thermal} for the dimensionful $\gamma$: 
$\gamma_D$ is the thermally averaged decay rate,
$\gamma_{N}^{\rm sub}$ represents $\Delta L=2$  interactions mediated by $N$.
As shown in~\cite{thermal}
there is no quasi-resonant enhancement of $\gamma_N^{\rm sub}$ at $T\sim m_{N_1}$
so that we can neglect it.
These approximations have been used to obtain eq.s\sys{gammaApprox}.
In our analytic approximations we neglect $\Delta L = 1$ scatterings, 
that at small temperature 
give corrections of relative order ${\cal O}(\lambda_t^2,g_2^2)/\pi^2$, e.g.}\\
\gamma_{Hs}(z\sim \infty)\approx\frac{3\MN^4}{16\sqrt{2}}
\lambdaN^2\lambda_t^2\frac{e^{-z}}{(\pi z)^{9/2}} &&
\gamma_{Ht}(z\sim \infty)\approx\frac{3\MN^4}{64\sqrt{2}\pi^2}
\lambdaN^2\lambda_t^2\frac{e^{-z}}{(\pi z)^{5/2}}\ln\frac{\MN}{m_h}\label{eq:mh}
\end{eqnsystem}
$\Delta L=1$ scatterings are subleading also at large temperature $T\gg\MN$,
because inclusion of thermal effects modify $\gamma_D\propto T^2$
into $\gamma_D \propto T^4$~\cite{thermal}.


\small 
\begin{thebibliography}{99}

\bibitem{thermal}
\hepart[hep-ph/0310123]{G. Giudice, A. Notari, M. Raidal, A. Riotto, A. Strumia}.

\bibitem{sun-exp}
\art{B.T. Cleveland}{\NP~(Proc. Suppl.)}{B38}{47}{1995};
\art[hep-ph/9812011]{SuperKamiokande collaboration}{Phys.Rev.Lett.}{82}{2430}{1999};
\art{K. Lande et al.}{Nucl.Phys.Proc.Suppl.}{77}{13}{1999};
\art{SAGE Collaboration}{Nucl.Phys.Proc.Suppl.}{77}{20}{1999};
\art{{\sc Gallex} collaboration}{Nucl.Phys.Proc.Suppl.}{77}{26}{1999}.


\bibitem{atm-exp} 
\art[hep-ex/9805006]{Super-Kamiokande collaboration}{\PL}{B436}{33}{1998};
D. Casper, talk presented at the XXXIV Recontres de Moriond on Electroweak
interactions and unified theories, Les Arcs, 13--20/3/1999,
available at the www address 
{\tt moriond.in2p3.fr/EW/transparencies.}

\bibitem{LSND}
\art{LSND collaboration}{\PR}{C54}{2685}{1996}
and
{\em Phys. Rev.} {\bf C58} (1998) 2489.

\bibitem{original} \art{M. Fukugita and T. Yanagida}{\PL}{B174}{45}{1986}.


\bibitem{nucleos} See, for example, \hepart[astro-ph/9905320]{K.A. Olive, G. Steigman and T.P. Walker}.


\bibitem{M>Trh}
For alternatives, see
\art[hep-ph/9905242]{G.F.~Giudice, M.~Peloso, A.~Riotto and I.~Tkachev}{\JHEP}{8}{14}{1999};
\hepart[hep-ph/9907559]{T.~Asaka, K.~Hamaguchi, M.~Kawasaki and T.~Yanagida}.

\bibitem{mucalcoli} \art{J.A. Harvey and M. Turner}{\PR}{D42}{3344}{1990};
\art{B.A. Campbell, S. Davidson, J. Ellis and K.A. Olive}{\PL}{B297}{118}{1992};
\art[hep-ph/9207221]{H. Dreiner and G.G. Ross}{\NP}{B410}{188}{1993}; 
\art[hep-ph/9401234]{S.A. Abel and K.E.C. Benson}{\PL}{B335}{179}{1994}.


\bibitem{olive} \art[hep-ph/9401208]{J.M. Cline, K. Kainulainen and K.A. Olive}{\PR}{D49}{6394}{1994}.


\bibitem{sphaleronOld}
\art{G. t'Hooft}{\PRL}{37}{8}{1976};
\art{V. Kuzmin, V.A. Rubakov and M.E. Shaposhnikov}{\PL}{155B}{36}{1985};
\art{J. Ambjýrn, T. Askgaard, H. Porter and M.E. Shaposhnikov}{\NP}{B353}{346}{1991}.

\bibitem{sphaleronNew}
It has been pointed out in
\art[hep-ph/9609481]{P. Arnold, D. Son and L.G. Yaffe}{\PR}{D55}{6264}{1997}
that hot sphaleron rates are suppressed by an extra factor $\alpha_2$
with respect to a `classical' estimate.
Lattice simulations confirm this extra factor: see
\hepart[hep-lat/9909054]{D. Bodeker, G.D. Moore, K. Rummukainen}.


\bibitem{epsN}
\art[hep-ph/9712468]{E. Roulet, L. Covi and F. Vissani}{\PL}{B424}{101}{1998};
\art[hep-ph/9710460 version 2]{W. Buchm\"uller and M. Pl\"umacher}{\PL}{B431}{354}{1998};
\art[hep-ph/9805427]{M. Flanz and E.A. Paschos}{\PR}{D58}{11309}{1998};
\art[hep-ph/9812256]{A. Pilaftsis}{Int. J. Mod. Phys.}{A14}{1811}{1999} (and references therein).



\bibitem{rho}
\art{A. Dolgov}{Sov. J. Nucl. Phys}{33}{700}{1981};
\art{L. Stodolsky}{\PR}{D36}{2273}{1987};
\art{G. Raffelt, G. Sigl, and L. Stodolsky}{\PRL}{70}{2363}{1993};
\art{G. Sigl and G. Raffelt}{\NP}{B406}{423}{1993}.


\bibitem{L-SM} \art[hep-ph/9604229]{M. Pl\"umacher}{Z. Phys.}{C74}{549}{1997}.
See also \art{M.A. Luty}{\PR}{D45}{455}{1992}.

\bibitem{L-MSSM}
\art[hep-ph/9704231]{M. Pl\"umacher}{\NP}{B530}{207}{1998}.



\bibitem{altri studi}
For related studies, see
\art[hep-ph/9810308]{Y. Buchm\"uller and T. Yanagida}{\PL}{B445}{399}{1999};
\hepart[hep-ph/9909477]{H. Goldberg}.

\bibitem{DI} \hepart[hep-ph/0202239]{S. Davidson and A. Ibarra}.

\bibitem{U1}
\art[hep-ph/9807228]{J.K.~Elwood, N.~Irges and P.~Ramond}{\PRL}{81}{5064}{1998};
\hepart[hep-ph/9808489]{P.~Ramond};
\art[hep-ph/9901243]{Q. Shafi and Z. Tavartkiladze}{\PL}{B451}{129}{1999};
\art[hep-ph/9902283]{S. Lola and G.G. Ross}{\NP}{B553}{81}{1999};
\hepart[hep-ph/9907532]{G. Altarelli and F. Feruglio}.


\bibitem{U2} \art[hep-ph/9903460]{R. Barbieri, P. Creminelli and A. Romanino}{\NP}{B559}{17}{1999}.

\bibitem{U2-BHRR}
\art[hep-ph/9610449]{R.~Barbieri, L.J.~Hall, S.~Raby and A.~Romanino}{\NP}{B493}{3}{1997}.


\bibitem{Kolb} See the discussion about GUT baryogenesis in `The early universe', E.W. Kolb and M.S. Turner,
Addison-Wesley ed.
See also  \art{E.W. Kolb and S. Wolfram}{\NP}{B172}{224}{1980}, erratum ibid. {\bf B195} (1982) 542.

\bibitem{Mathematica}
S. Wolfram, {\em The Mathematica book}, 3rd ed.
(Wolfram Media/Cambridge University Press, 1996).


\bibitem{Plum2}
\art[hep-ph/0205349]{W.~Buchmuller, P.~Di Bari, M.~Plumacher}{\NP}{B643}{367}{2002}.

\bibitem{SMRGE}
P.~H.~Chankowski and Z.~Pluciennik,
{\em Phys.\ Lett.}  {\bf B316} (1993) 312
({\em hep-ph/9306333});
K.~S.~Babu, C.~N.~Leung and J.~Pantaleone,
{\em Phys.\ Lett.}  {\bf B319} (1993) 191
({\em hep-ph/9309223}).
An error was corrected in
S.~Antusch, M.~Drees, J.~Kersten, M.~Lindner and M.~Ratz,
{\em Phys.\ Lett.}  {\bf B519} (2001) 238
({\em hep-ph/0108005}).

\end{thebibliography}
\end{document}